\documentstyle[12pt,epsfig]{article}

\newlength{\dinwidth}
\newlength{\dinmargin}
\def\lapproxeq{\lower .7ex\hbox{$\;\stackrel{\textstyle
<}{\sim}\;$}}
\def\gapproxeq{\lower .7ex\hbox{$\;\stackrel{\textstyle
>}{\sim}\;$}}

\def\be{\begin{equation}}
\def\ee{\end{equation}}
\def\bea{\begin{eqnarray}}
\def\eea{\end{eqnarray}}

\def\lapproxeq{\lower .7ex\hbox{$\;\stackrel{\textstyle
<}{\sim}\;$}}
\def\gapproxeq{\lower .7ex\hbox{$\;\stackrel{\textstyle
>}{\sim}\;$}}
\def\be{\begin{equation}}
\def\ee{\end{equation}}
\def\bea{\begin{eqnarray}}
\def\eea{\end{eqnarray}}

\begin{document}
\titlepage

\vspace*{2cm}

\begin{center}
{\Large \bf High-energy cosmic gamma rays \\ from the
`Single Source'\rule{0mm}{3ex}}

\vspace*{1cm} \textsc{A.D. Erlykin$^{a,b}$ and A.W.
Wolfendale$^{b}$}

\vspace*{0.5cm} $^a$ Lebedev Physical Institute, Leninsky Prospekt, Moscow, Russia\\
$^b$ Department of Physics, University of Durham, DH1 3LE, UK \\
\end{center}

\begin{abstract}
Some six years ago, we (Erlykin and
Wolfendale, 1997) proposed the `single source model' in which a
local, recent supernova remnant (SNR) was responsible for the
 `knee' in the cosmic ray (CR) energy spectrum at $\sim3$~PeV.
Stimulated by the paper by Bhadra (2002), which drew attention to
a possible gamma ray signature of this local remnant,
we now study the situation for the local source and we
conclude that, in contrast to Bhadra's conclusion, the
non-observation of this remnant is understandable - at least
using our SNR model. It is due to the fact that this SNR, being local, develops
in the local hot interstellar medium (~HISM~) with its low density of gas and
also being nearby it will be an extended source occupying up to
40$^\circ$ of the sky and thus indistinguishable from the background. 

\end{abstract}

\section{Introduction}

The `knee', a rather sharp steepening in the primary cosmic ray (~CR~)
energy spectrum at about 3 PeV, was inferred from the observation of a
similar feature in the measured size spectrum of extensive air showers
by Kulikov and Khristiansen~(1958). The knee is commonly
asserted to be due to an increasing failure of `Galactic containment' 
of the CR generated by sources within the Galaxy, the containment
being caused by the magnetic fields in the interstellar medium (~ISM~), 
However, it is the firmly-held view of the
present authors that the knee is too sharp for this explanation
and we have advanced what we claim to be a more realistic
model. This is our `single source (~SS~) model' (see Erlykin and
Wolfendale, 1997, 2001b for recent details) which comprises cosmic ray
acceleration up to the knee energy by supernova remnants, the knee
itself being due to the truncation that occurs at 3~PeV for oxygen
nuclei from a single, recent, nearby SNR. The other main
accelerated nucleus at these energies is iron and its termination
occurs at about 12~PeV where, it is claimed by us that there is a small
second knee (when the spectrum is plotted as $E^3I(E)$~vs.~$E$,
the knees appear as small peaks). The remainder of the CR spectrum
(~at least to some 10$^{9}$ GeV, or so~) is presumed due to `super'-SNR
and other sources and their spatial distribution is such as to give a
comparatively smooth spectrum in the PeV region.

Erlykin and Wolfendale (to be referred to henceforth as EW) have
examined a variety of other cosmic ray data and concluded that
there is either support for the model or that the data are
neutral. Very recently, low energy gamma ray data have also been
studied (EW, 2002) and the well-known `gamma ray excess' in the
Inner Galaxy, and deficit in the Outer Galaxy, have been explained
in terms of propagation differences dependent on the conditions in
the ISM from which the gamma rays come. The results relate to
Galaxy-wide properties and, although the SNR acceleration
hypothesis has been invoked, there is no significant information
about the single source.

It is at higher gamma ray energies where potential problems exist
(e.g. Drury et al., 1994). Most recently, Bhadra (2002) has argued
that the single source should be visible in TeV gamma rays, and it
is not. This is the topic to be addressed here. We use the results
found in a very recent paper (EW, 2003a to be referred to as~I),
where we made predictions of the fluxes and the angular distribution
of gamma ray
intensity from SNR of different ages and at different distances
from the Sun. The threshold energies were taken as 0.1~GeV and
1~TeV.

A critical feature of the Bhadra estimate was the `normalisation'
of the SNR `conditions' so as to give the CR energy density created
by the single source at earth. We regard this as a legitimate
procedure and we follow this path, although other features of our
model differ considerably from those adopted by Bhadra. Our
calculations are thus not simply a `re-run of the Bhadra calculations
with different values for the parameters' but, rather, for what is
certainly a more appropriate model of SNR acceleration and (~less
certainly, perhaps~) a significantly different model of cosmic ray
propagation.

\section{The Bhadra estimate}

Bhadra's model is rather straightforward, in principle, at least: 
particles are accelerated by
the SNR shock to give a differential spectrum $AE^{-\gamma}$, with
$\gamma = 2.0,$ and these particles interact with the ambient (or
swept-up) gas of density $n$, with $n=1$~cm$^{-3}$. At this stage it
is necessary to make critical remarks, however. Implicit in the Bhadra
calculations is the assumption that the SNR shock accelerates the CR
instantaneously, at `$t = 0$'. Although this can be used to give a 
viable mathematical model,
such a situation is certainly not appropriate to a real SNR where the
acceleration occurs over an extended period: $8\cdot 10^4$ years in
our model, and little different in other SNR acceleration models.

Nevertheless, we continue to describe the Bhadra calculations. It is
assumed that the particles diffuse from the source a distance $r$ from the Sun
in a normal, Gaussian fashion with diffusion coefficient $D=10^{29}$~cm$^2$s$^{-1}$. The
parameters are chosen to give the required CR energy density at the
Earth created by the single source. For instance, if the source is at
the distance of 300 pc and it is 10$^4$ years old the total energy
transferred from SN to CR is required to be 1.9$\cdot$10$^{50}$ erg.
The Table summarizes the most important parameters.

\begin{table}
\begin{center}
\begin{tabular}{l @{\qquad } c @{\qquad } c}
 & \raisebox{1ex}{Bhadra} & \raisebox{1ex}{EW} \\
$E_{SN,CR}$\ ($10^{50}$~erg) & 1.9 & 1.0 \\
$D$\ ($10^{29}$~cm$^2$s$^{-1}$) & 1.0 & $2.25\left(E/10^3\right)^{0.5}$ \\
$n$\ (cm$^{-3}$) & 1.0 & $3\times10^{-3}$ \\
$\gamma_p$ & 2.0 & 2.15 \\
$F_\gamma(>0.1\:{\rm GeV})$~(cm$^{-2}$s$^{-1}$) & $2\times 10^{-7}$
& $10^{-8}$ \\
$F_\gamma(>1\:{\rm TeV})$~(cm$^{-2}$s$^{-1}$) & $4\times 10^{-11}$
& $0.7\times 10^{-12}$ \\
\end{tabular}
\end{center}
\caption { Values of the parameters adopted by Bhadra (2002) in
comparison with those in the present work (denoted EW). The
remnant is taken to be at 300~pc from the Sun. 
$E_{SN,CR}$ is the cosmic ray energy input from the SNR;
$D$ is the diffusion coefficient for normal, gaussian diffusion, $E$ is in
GeV; $n$ is the density of the interstellar medium (~ISM~) in {\em
H}-atoms cm$^{-3}$; 
$\gamma_p$ is the exponent of the differential proton spectrum;
$F_{\gamma} ( > 0.1 GeV )$ and $F_{\gamma} ( > 1 TeV )$ are the predicted
gamma ray fluxes.} 
\end{table}

The expected minimum gamma ray flux above 0.1 GeV rises with the age
of the SN  from
$0.2\cdot 10^{-7} {\rm cm}^{-2}{\rm s}^{-1}$ at 10$^3$ years to
$2\cdot 10^{-7} {\rm cm}^{-2}{\rm s}^{-1}$ at 10$^5$ years and from
$0.4\cdot 10^{-11} {\rm cm}^{-2}{\rm s}^{-1}$ to
$4\cdot 10^{-11} {\rm cm}^{-2}{\rm s}^{-1}$ above 1 TeV,
respectively. For the real candidates for the single source discussed
in EW, (1997) (~eg. Loop I, Clayton SNR~) the expected fluxes are
substantially higher. Comparing these fluxes with the diffuse gamma
ray background Bhadra found that for the present gamma ray
telescopes it should have been possible to observe the single
source. Since there
has been no claim for an observation Bhadra concluded that the single
source cannot be such a SNR.

Turning to our remarks on the validity of Bhadra's model, in addition
to the basic problem with the assumption about the
instantaneous acceleration there are two further reasons why we cannot allow 
this conclusion to stand and,
in fact, Bhadra made the neeeded reservations, viz. `{\em the
detection could be crucial, depending on the angular size of the object'} and
`{\em unless the source is in a lower density environment'}. These are
the points addressed in the following sections. We require, first, the
likely whereabouts of the single source such that it
can give the particle spectrum needed in the knee region.

\section{The present treatment}

\subsection{The predicted particle spectrum}

We have calculated cosmic ray energy spectra originating from SNR of
different ages and distances. The model of acceleration was described
in EW (2001a) and briefly in I. The propagation of accelerated cosmic
rays through the ISM was calculated using two
alternative assumptions about the mode of propagation from the source: 
`anomalous' diffusion, viz. making allowance for the fractal-like nature
of the ISM, and normal, Gaussian, diffusion. Following the work of Lagutin et
al.(2001a,b) we distinguish these two modes by a parameter $\alpha$
determined by the fractal nature of the medium; $\alpha$ is equal to 1
or 2 for anomalous or normal diffusion, respectively. 

The difference between the two modes can be seen most clearly in the
shape of the
lateral distribution function for the cosmic ray intensity: 
$\frac{1}{(1+x^2)^2}$ for $\alpha = 1$ and $exp(-\frac{x^2}{4})$ for
$\alpha = 2$, with $x = \frac{r}{R_d}$, $r$ being the distance from
the radius $R_s = 100$ pc where the particles start to diffuse and
$R_d$ being the diffusion radius which is defined as $R_d =
H_z(\frac{t}{\tau(E)})^{\frac{1}{\alpha}}$, i.e., there is a different
time dependence for the two modes. $H_z = 1$ kpc for the vertical scale
of the galactic halo and $t$ and $\tau(E)$ are the diffusion time and
time against escape, respectively, for the protons. Details of the propagation
model are given in EW (2002) and in I. 

The difference in the two lateral distributions is quite
dramatic. Thus, for $x =$ 2, 4, 6 and 8, the ratio of the intensity
for anomalous diffusion to that for normal diffusion changes as 0.11,
0.19, 5.9 and 2.1$\cdot$10$^3$, respectively. The long tail for
anomalous diffusion - the occasional considerable `penetration'
(~'Levy flights'~) - can have important implications. An example is
that for the secondary to primary ratio; to our knowledge the
implications have not been worked out.

We have determined the proton energy spectra expected for the two
values of $\alpha$ and these are given in Figures 1 and 2.  
Also indicated in the Figures is the spectrum `required' by the SS
model. We have argued that the `needed' particles are oxygen nuclei
for the first `peak' at 3~PeV and iron nuclei for the second `peak' at
$\sim12$~PeV, and the requirement has been converted to rigidity
before plotting, in Figures~1 and 2. An alternative association of
the `peaks' with helium and oxygen changes the indicated SS spectrum no
more than by 17\%. (~EW, 2003b~).

Dotted lines above and below the `SS' curve indicate its
uncertainty limits; the least uncertainty and therefore the most
important constraint of the SS model is in the knee region. The limits
come
from the fit to the experimental data on the primary energy spectrum
measured by means of the Cherenkov light emitted by extensive air showers
(~EW, 2001b~). These data determine the magnitude of the uncertainty
in the range of about a decade below the knee. At lower energies, - 
$10^3 - 10^4$ GeV - the upper limit is determined by the uncertainty of
the direct measurements of the primary CR energy spectrum (~Biermann and 
Wiebel-Sooth, 1999~). Although the shape of the SS spectrum has been 
adopted from the theoretical model of Berezhko et al.(~1996~) its lower
experimental limit at low energies is completely uncertain because the 
contribution of the Single Source to the total CR intensity at these
energies is negligibly small and even consistent with zero. However, the
actual value of the uncertainty at TeV energies is not important for this
analysis because we have made a quantitative comparison of the calculated
proton spectra with the SS spectrum only in the most important range of
energies, viz. that covering a decade below the knee. The energy range
of $logE =
4.6 - 5.6$ used for this comparison is indicated in the Figures by {\em `min'} and {\em `max'}.

\begin{figure}[htbp]
\begin{center}
\includegraphics[height=14cm,width=16cm]{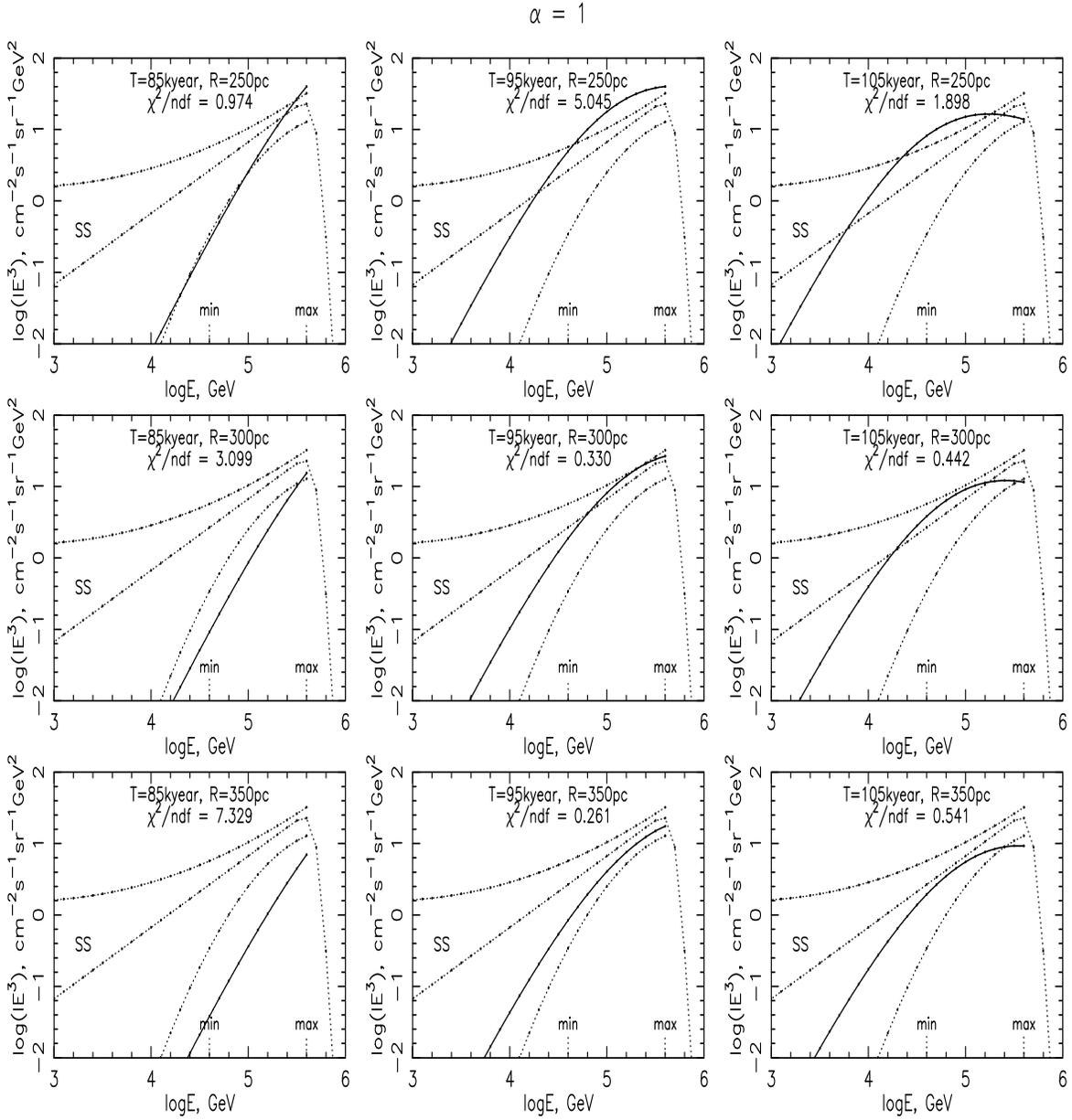}
\vspace{1cm}
\caption{\footnotesize Energy spectra of protons (~full lines~)
expected from SNR of
different ages (T) and at different distances (R) from the Sun,
as indicated inside the graphs, compared with the rigidity
spectrum of CR according to our Single Source Model (~dotted line
denoted by SS~). Dotted lines above and below the SS line indicate its
uncertainty limits. SNR protons propagate through
the ISM by means of anomalous diffusion with $\alpha = 1$ (~EW,
2002~) (~rigidity and energy are, of course, the same for protons~).
The energy interval used for the comparison of proton spectra with the
SS model is marked by {\em min} and {\em max}. The result of the
comparison in terms of the reduced $\chi^2$, i.e. $\chi^2$ divided by
the number of degrees of freedom, $\chi^2/ndf$, is shown inside the Figures.}
\end{center}
\label{fig:bhad1}
\end{figure}

\begin{figure}[htbp]
\begin{center}
\includegraphics[height=14cm,width=16cm]{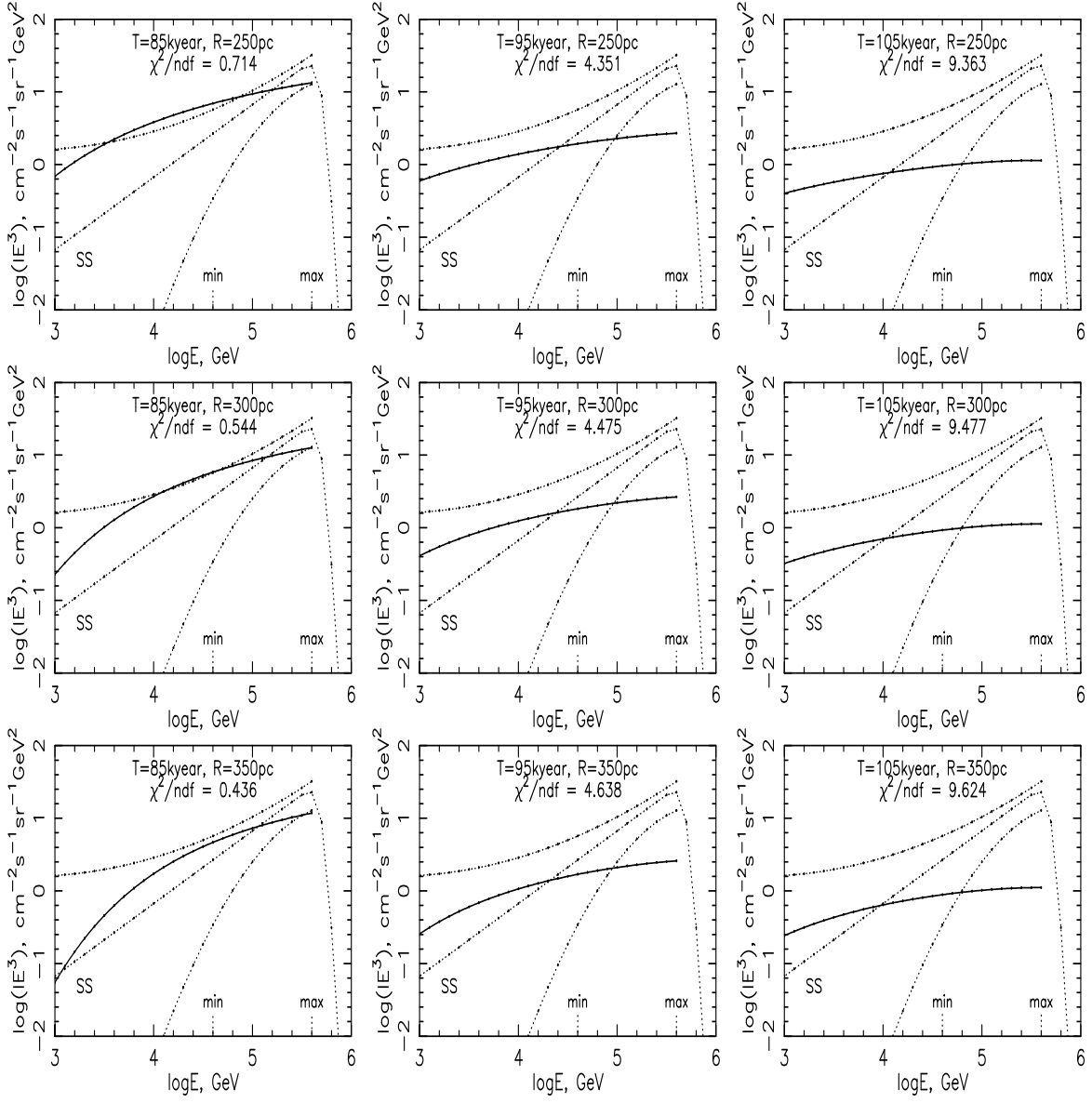}
\vspace{1cm}
\caption{\footnotesize The same as in Figure 1, but for normal
diffusion with $\alpha = 2$.}
\end{center}
\label{fig:bhad2}
\end{figure}
Some comments are needed about the Figures. It is evident that
$\alpha=2$ always gives a bad fit (Figure~2) whereas $\alpha=1$
can give a reasonable one, at least for energies above 10$^4$ GeV, up
to the cut-off at 4$\cdot10^5$ GeV, for a range of age (T) and distance (R)
values \footnote{In spite of the fact that such an integral characteristic as
$\chi^2/ndf$ has a formally acceptable value for $\alpha = 2$ at T$\approx 85$
kyear the slopes of the proton spectra are too steep to give the sharp
knee and we do not consider it as a good contender.}. 
Calculations made for a wide range of T and R allow us to
estimate the range over which there is satisfactory agreement between 
calculations
and the SS model; it is $85<T<115$ kyear and $250<R<400$
pc for the adopted set of input parameters.

It is relevant to point out that larger distances, too, would give
a good fit to the spectral {\em shape}. The necessary upward
movement in intensity could be effected by increasing the fraction
of the shock energy going into CR. As was pointed out in~I,
Berezhko et~al.,(1996) used much bigger values than the 10\% used here
- their highest being 80\%. Specifically, for 100~kyear an increase in CR yield
by a factor 8 would allow the source to be at $\sim 650$~pc.

The shape and the absolute intensity of the CR energy spectrum give the most
stringent constraints on the age and distance. The T-R region of SNR
which could give an acceptable spectrum (~for our standard `10\%'~) is
shown in Figure 3 by the
95\% confidence level contour. Following Bhadra's approach
we have also used such an integrated characteristic of
the spectrum as the energy density contained in it. Again, we used for
the comparison just the last decade of the
spectrum below the knee, because intensities at low energies are
so poorly determined by
the Single Source Model. The energy density contained in the spectrum
of our single source between $4\cdot 10^4$ GeV and $4\cdot 10^5$ GeV is
1.84$\cdot 10^{-4} {\rm eV}{\rm cm}^{-3}$. Comparison of this value,
allowing for uncertainty, with those expected for SNR of different ages
and distances gives an acceptable T-R region indicated by the dashed line
in Figure 3. It overlaps with the region deduced from the comparison
of the spectral shapes. This proves the consistency of these results,
although we must admit that the two methods are not completely independent,
because consistent values for the spectra should inevitably give
consistent 
energy densities contained in them.

In any case this analysis indicates that our single source should be located
at about 300 - 350 pc from the Sun and should be about 90 - 100 kyears old. We
know of no objection to such parameters (~see \S4 for its likely location~).
\begin{figure}[htb]
\begin{center}
\includegraphics[height=7cm,width=15cm]{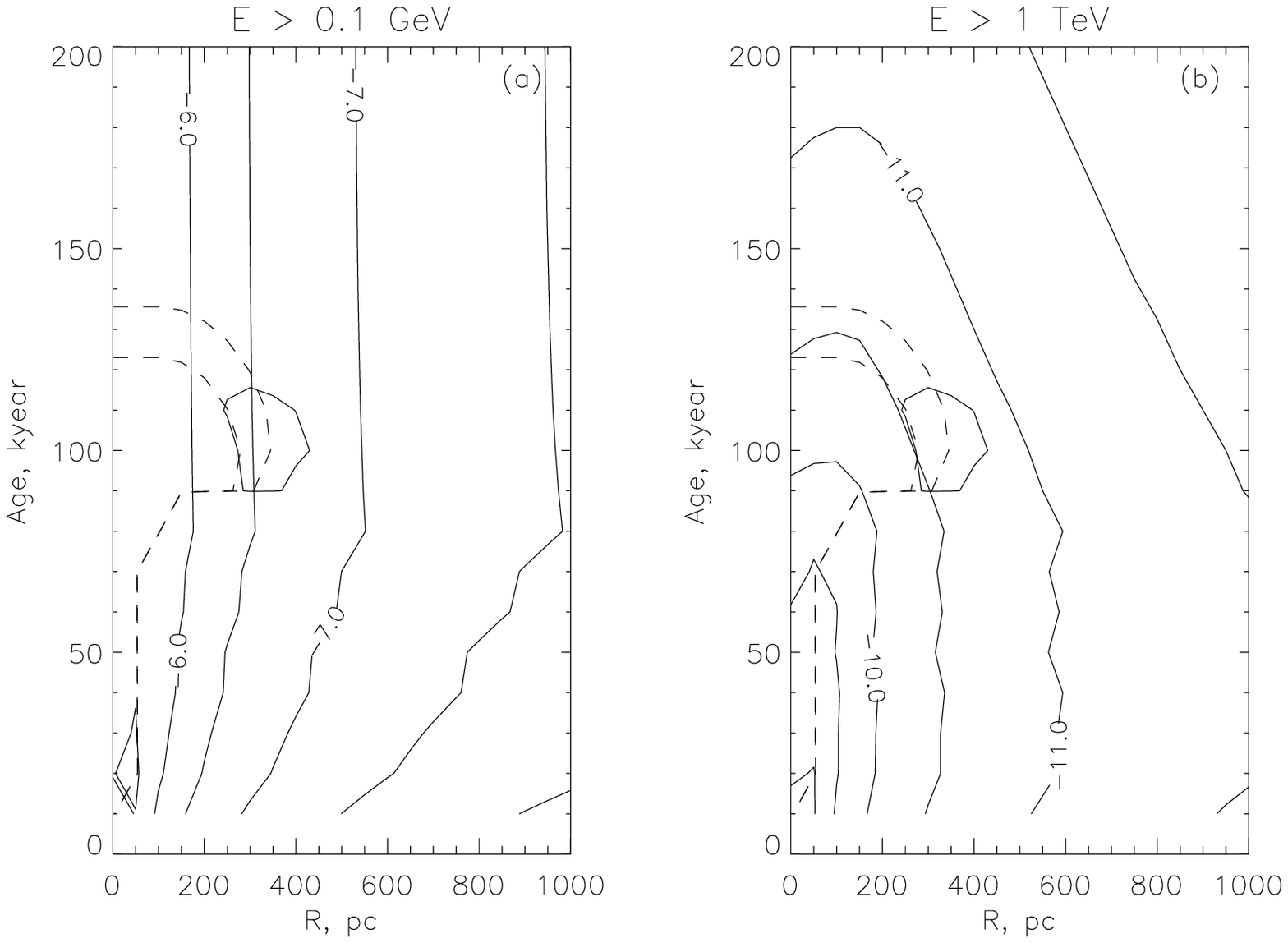}
\caption{\footnotesize Age - Distance diagram for a SNR which is
responsible for the formation of the knee - i.e. for our 'Single
Source'. The inserted full-line contour in the center indicates the
95\% confidence level region deduced from the analysis of the shape
and the intensity of the CR particle spectra; the dashed lines show limits
based on the comparison of the energy density contained in the Single
Source. Full-line contours show the expected total gamma ray fluxes
with the energy of gamma quanta above 0.1 GeV (a) and 1 TeV (b)
expected for PP-interactions. Labels on the contours indicate values of
log(Flux), the flux being in units of cm$^{-2}$s$^{-1}$. }
\end{center}
\label{fig:bhad3}
\end{figure}
\subsection{The predicted gamma ray flux}

\subsubsection{Calculations for the standard gas density}

Fluxes of gamma rays from SNR of different ages and at different
distances from the Sun have been calculated in I.
In Figure 3 we show contours of the fluxes calculated
for the interactions of protons, accelerated by the SNR, which propagate
through the ISM by way of anomalous diffusion (~$\alpha = 1$~) and
collide with nuclei of the ISM, The density of the ISM for these
calculations has been taken to be our standard $n = 1 {\rm cm}^{-3}$. It is
seen that the expected fluxes are about 3$\cdot 10^{-7}{\rm
cm}^{-2}{\rm s}^{-1}$ for $E_\gamma > 0.1$ GeV and about 2$\cdot 10^{-11}{\rm
cm}^{-2}{\rm s}^{-1}$ for $E_\gamma > 1$ TeV.

It is necessary now to study the implications of the fact that the
cosmic rays from our Single Sourse are not only protons, but that the
composition is mixed. At the same rigidity it consists of 21\%
P, 48\% O, 13\% `heavy' nuclei ($10<Z<23$) and 18\% Fe. Because nuclei
of the same rigidity are more efficient in the production of gamma quanta
(~see I~), the expected fluxes should be increased by a factor of
10.5. This gives expected fluxes of $\sim 3\cdot 10^{-6}{\rm
cm}^{-2}{\rm s}^{-1}$ for $E_\gamma > 0.1$ GeV and $\sim 2\cdot 10^{-10}{\rm
cm}^{-2}{\rm s}^{-1}$ for $E_\gamma > 1$ TeV. It is necessary to point
out that the CR energy injection is higher by the same amount.

\subsubsection{The appropriate gas density to adopt}

Since this is a first order effect, a detailed examination of the
problem is required. As pointed out in I, there are several aspects
that need attention, specifically: \\
(i) the effect of the pre-SNR winds from the progenitor star in
excavating a `hole' into which the SNR shock expands, this is
essentially the origin of the HISM (~note that for vey young SNR,
however, the progenitor star's wind may have {\em enhanced} the density~); \\
(ii) the general ISM density in the region where the progenitor star
was situated and \\
(iii) the pressure, or otherwise, of clumpy gas (~often molecular~) in
the vicinity of the SNR.

In the present case, it is almost certain that the progenitor star -
at 300-350 pc from the sun - was in the Hot Interstellar Medium, where the
density is often quoted as $n \simeq 3\cdot 10^{-3}$ cm$^{-3}$
(~Berezhko et al., 1996; Cox and Reynolds, 1987, for example quote a
density of $4\cdot 10^{-3}$cm$^{-3}$~). Indeed, since all our model 
calculations assume this to be the case, a low density target material is a
prerequisite. The assumption of the HISM comes from two factors: \\
(i) the absolute maximum particle energy needed to explain the knee
(~$3\cdot 10^6$ GeV for oxygen nuclei~) appears only for this density,
in the Berezhko et al. model, and \\
(ii) the HISM is eminently reasonable for a nearby source.

Factor (ii) can be examined in more detail. Insofar as the sun is
located on the edge of a spiral arm, in one hemisphere (~the South~),
at least, the gas density will be low. Frish (~1997~) has considered
the situation in detail. Beyond the very local region, where there is
the `Local Fluff' (~of extent $\sim$3 pc~) in certain directions the
column density of atomic (~and molecular~) hydrogen is very low. The
`Local Bubble' (~Loop I~) is quickly reached, this Bubble being caused
by several SN over the past Myear. Here, one expects the HISM, with
its density $\sim 3\cdot 10^{-3}$ cm$^{-3}$. Concerning the interarm
region, Frish quotes an
undisturbed part as having a density of $4\cdot 10^{-4}$ cm$^{-3}$,
i.e. even smaller. However, much of the Local Bubble is here and the
density will be higher because of material brought in from elsewhere.

Interestingly, Frish (1981) suggested, earlier, that the sun is
embedded in one of the super-bubble shells associated with the
formation of the Scorpius-Centaurus Association. Another possibility
for the formation of Loop I (`but very similar to that given above is
that it was caused by activity in the `Upper Centaurus Lupus' subgroup
some 14-15 Myear ago (~De Geus, 1991~)).

A distinction must be made between a `recent' SN exploding in Loop I,
i.e. in the low density HISM caused by previous SNR and stellar
activity, and the low level gamma ray flux from Loop I as a
whole. This latter was considered in I and it was argued there that
the mean density overall, allowing for molecular clouds inside the
Loop, and the piled-up gas in the edges of the Loop, is $\langle n
\rangle \simeq$ 0.1 cm$^{-3}$. There is no conflict with our $n =
3\cdot 10^{-3}$ cm$^{-3}$ if the single source is well into the interior.

\subsubsection{The predicted gamma ray flux}

With $n = 3\cdot 10^{-3}$, the predicted flux is reduced to $\sim
10^{-8}$ cm$^{-2}$s$^{-1}$ for $E_\gamma > $0.1 GeV and $\sim
0.7\cdot 10^{-12}$cm$^{-2}$s$^{-1}$ for $E_\gamma >$ 1 TeV.

If, surprisingly, the source is not in the HISM in the Local Bubble, but is
isolated, then as was discussed in I, $n\sim0.1{\rm cm}^{-3}$ and
the predicted fluxes are $\sim 3\cdot10^{-7}{\rm cm}^{-2}{\rm s}^{-1}$ 
for $E_\gamma > 0.1$ GeV and 
$\sim 2\cdot 10^{-11}{\rm cm}^{-2}{\rm s}^{-1}$ for $E_\gamma > 1$ TeV.

The predicted fluxes are shown in Figure 4, for the various
possibilities of {\em n} and {\em Z}.
\begin{figure}[htb]
\begin{center}
\includegraphics[height=9cm,width=15cm]{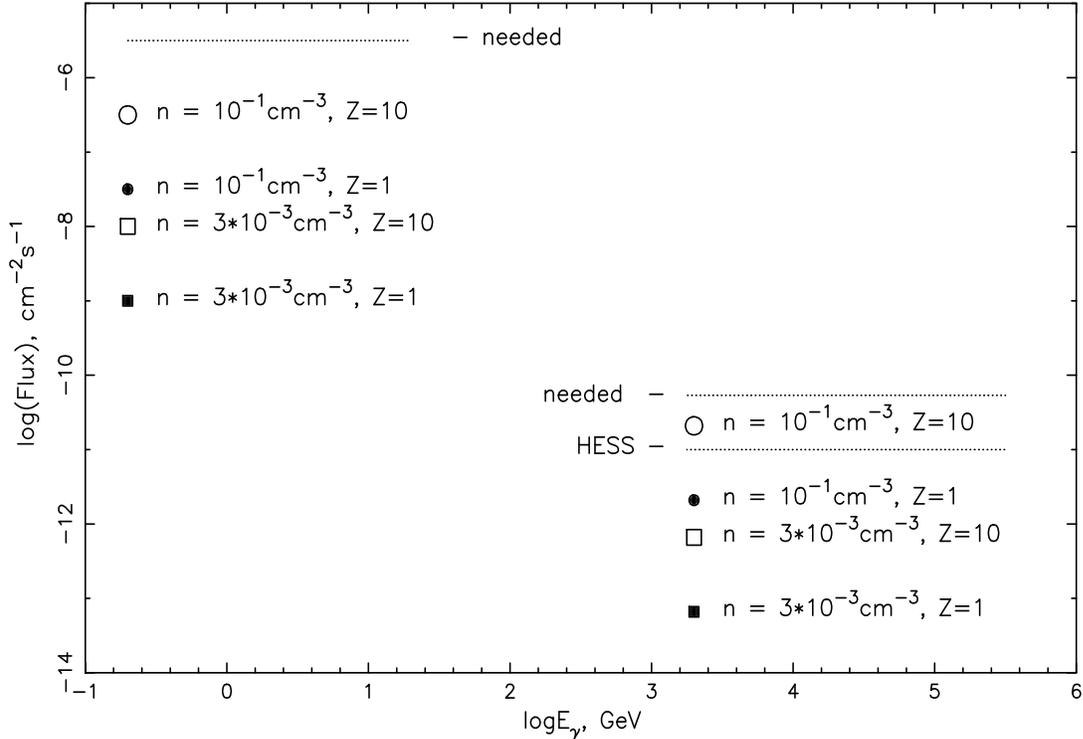}
\vspace{2mm}
\caption{\footnotesize Schematic representation of the gamma ray fluxes
from the claimed `Single Source' at energies $E_\gamma > 0.1$ GeV and $> 1$
TeV for a variety of scenarios. The source would be at $\sim 300-350$ pc
from us and have an age of
$\sim 90-100$ kyear, i.e. it would be only 10-20 kyear since the particles
were released by the remnant. The dotted lines marked `needed' are the 
minimum fluxes
required for the source to have been detected by the arrays in
use up to now. As an
indication of the future we also give the estimated minimum detectable
flux for $E_\gamma > 1$ TeV for a source of 20$^\circ$ radius, from the
work of Aharonian et al.,(1997). These authors give results for 1000
hours of observation of a 'point' source and one of 1$^\circ$ extent;
our estimate arises from an extrapolation, based on results described
in I. The work
reported relates to the then proposed IACT (~100 GeV - class imaging
atmospheric telescope array~), denoted `HESS'.}
\end{center}
\label{fig:bhad4}
\end{figure}
 
\subsection{The limiting sensitivity in practice}

The sources must be detected against a background due to the
Galactic diffuse emission arising from CR - ISM nucleus
interactions (~and, in the case of TeV gamma rays, protons interacting
in the atmosphere~). This background can be allowed for in a straightforward
manner for `point' sources, where a subtraction can be made of the
signal nearby to the source, but for extended sources the
problem is much more severe. 
The angular radius of a SNR, which is 90-100 kyear old, seen from a
distance of 300-350 pc is about 20$^\circ$ (~see also Figure 5 in I~).
For such a large size the determination of the background by a linear
interpolation of the intensity between the adjacent regions is not at
all accurate. Estimates of the background based on the known
column density of the target gas, although more appropriate, are again
not sufficiently accurate because the intensity of the initiating cosmic
rays along the line of sight cannot be assumed to be strictly constant.

Estimates of the limiting fluxes were made from the available data and
they are given in I. For a source of radius 20$^\circ$ the average limits are
$3\cdot 10^{-6} {\rm cm}^{-2}{\rm s}^{-1}$ for $E_\gamma > 0.1$ GeV
and $5\cdot 10^{-11} {\rm cm}^{-2}{\rm s}^{-1}$ for $E_\gamma > 1$
TeV. These limiting sensitivities, which relate to observations made
so far, are indicated in Figure 4.

The lower energy limit comes from an extrapolation of the EGRET
results (~Hartman et al, 1999~), which refer to small angular sizes,
to 20$^\circ$. That for the upper energy band comes largely from the
Tibet and HEGRA arrays (~Amenomori et al., 2001 and Lampeitl et al., 
2001, respectively~).

It is instructive to examine these sensitivities in terms of the
background fluxes. For $E_\gamma > 0.1$ GeV, the intensity at low
latitudes towards the Galactic Anticenter is about 
$10^{-4} {\rm cm}^{-2}{\rm s}^{-1}{\rm sr}^{-1}$ and for our
20$^\circ$-radius source (~$\Omega$ = 0.4 sr~) the flux would be 
$4\cdot 10^{-5} {\rm cm}^{-2}{\rm s}^{-1}$. The plotted limit of 
$3\cdot 10^{-6}{\rm cm}^{-2}{\rm s}^{-1}$ is thus 8\% of the
background. For the direction to the Galactic Center, where the background
is higher by a factor of 3, this percentage is reduced to (2-3)\%.
When allowance is made for other sources of uncertainty our adopted
limit seems reasonable.

For $E_\gamma > 1$ TeV, the corresponding background in the Galactic
Anticenter direction has been
estimated to be $\sim 1.6\cdot 10^{-9}{\rm cm}^{-2}{\rm s}^{-1}$ 
(~Porter and Protheroe, 1999~), viz a flux of 
$(6-7)\cdot 10^{-10} {\rm cm}^{-2}{\rm s}^{-1}$ over a
20$^\circ$-radius source. Our adopted limit of 
$5\cdot 10^{-11}{\rm cm}^{-2}{\rm s}^{-1}$ is thus the same: $\sim$8\%
of the background. In the direction of the Galactic Center the
background is expected to be twice as large and the fraction of the
limiting flux falls to $\sim$4\%. This figure, too, is reasonable
in view of misidentified proton contributions and a variety of
technical problems associated with making absolute measurements at
different zenith angles. 

\section{Conclusions}

The conclusions are indicated in Figure~4. Only if the single
source is in a `high' density region ($n\sim0.1\ {\rm cm}^{-3}$) and the
primaries are `heavy' nuclei (~Z = 10~) (~with consequent high CR
energy injection~) will it
be possible to detect the Single Source at high energies. Such a
situation is not impossible but there would be problems for our Single
Source model with such a high ambient density, specifically that it
would not be possible with the model in use to reach the required 3
PeV energy at the knee. Insofar as we consider it very likely that the 
source is in the HISM and is extended, the chance of
detecting it with contemporary instruments is considered, by us, to be
very low. There are, however, hopes for the future. It is germane to
consider in which direction future, improved, gamma ray detectors
should be pointed in order to see it. The best that can be done here
is to suggest the general direction of Loop I. However, since this
structure occupies about 25\% of the sky the `advice' is not very helpful.

\section*{Acknowledgments}

The authors are grateful to The Royal Society and The University of
Durham for financial support. Dr. Paula Chadwick, of the HESS
collaboration, is thanked for providing useful information. Two
unknown referees are also thanked for useful comments and suggestions.

\section*{References}

\noindent Aharonian, F.A. et al., 1997, {\em Astroparticle Physics},
{\bf 6}, 369 \\
\noindent Amenomori, M. et al., 2001, {\em 27th Int. Cosm. Ray Conf.} 
(~Hamburg~), {\bf 6}, 2344 \\
\noindent Berezhko, E.G., Elshin, V.K. and Ksenofontov, L.T.,
1996, {\em J. Exp. Theor. Phys.}, {\bf 82}, 1 \\
\noindent Bhadra, A., 2002, {\em J. Phys. G: Nucl. Part. Phys.}, {\bf 28}, 397 \\
\noindent Biermann P.L. and Wiebel-Sooth B., 1999, {\em Astronomy \&
Astrophysics - Interstellar Matter, Galaxy, Universe},
Landolt-B\'{o}rnstein, Springer Verlag, Berlin/Heidelberg, {\bf 3}, 37 \\
\noindent Cox D.P. and Reynolds R.J., 1987, {\em Annual Rev. Astron. 
Astrophys.} {\bf 25}, 303 \\
\noindent De Geus, E.J., 1991, {\em Astron. Astrophys.}, {\bf 262},
258 \\
\noindent Drury, L.O'C., Aharonian, F.A. and V\"{o}lk, 1994, {\em
Astron. Astrophys.}, {\bf 287}, 959 \\
\noindent Erlykin, A.D. and Wolfendale, A.W., 1997, {\em J. Phys. G:
Nucl. Part. Phys.}, {\bf 23}, 979 \\
\noindent Erlykin, A.D. and Wolfendale, A.W., 2001a, {\em J. Phys. G:
Nucl. Part. Phys.}, {\bf 27}, 941 \\
\noindent Erlykin, A.D. and Wolfendale, A.W., 2001b, {\em J. Phys. G:
Nucl. Part. Phys.}, {\bf 27}, 1005 \\
\noindent Erlykin, A.D. and Wolfendale, A.W., 2002, {\em J. Phys. G:
Nucl. Part. Phys.}, {\bf 28}, 2329  \\
\noindent Erlykin, A.D. and Wolfendale, A.W., 2003a (~submitted to
J. Phys. G~) \\
\noindent Erlykin, A.D. and Wolfendale, A.W., 2003b, {\em
Nucl. Phys. B (~Proc. Suppl.~)} (~in press~) \\
\noindent Frisch, P.C., 1981, {\em Nature}, {\bf 293}, 377 \\
\noindent Frisch, P.C., 1997, astro-ph/9705231 \\
\noindent Hartman, R.C. et al., 1999, {\em
Astron. Astrophys. Suppl. Ser.},
{\bf 123}, 179 \\
\noindent Kulikov, G.V. and Khristiansen, G.B., 1958, {\em
J. Exp. Theor. Phys.}, {\bf 35}, 635 \\
\noindent Lagutin, A.A. et al. 2001a, {\em Nucl. Phys. B
(~Proc. Suppl.~)}, {\bf 97}, 267 \\
\noindent Lagutin, A.A. et al. 2001b, {\em Proc. 27th Int. Cosm. Ray
Conf.} (~Hamburg~), {\bf 5}, 1900 \\
\noindent Lampeitl, H. et al., 2001, {\em 27th Int. Cosm. Ray Conf.}
(~Hamburg~), {\bf 6}, 2348 \\
\noindent Porter T.A. and Protheroe R.J., 1999, {\em 26th
Int. Cosm. Ray Conf.}, Salt Lake City, {\bf 4}. 306 \\

\end{document}